\documentclass{ifacconf}

\usepackage{graphicx}      
\usepackage{natbib}        

\usepackage{amsmath}
\usepackage{amsfonts}
\usepackage{amssymb}
\usepackage{multirow}

\usepackage{xcolor}

\usepackage{standalone}

\usepackage{subcaption}

\usepackage{tikz}
\usepackage{pgfplots}
\newcommand{\figref}[1]{\figurename~\ref{#1}}
\newcommand{\tabref}[1]{\tablename~\ref{#1}}
\newcommand{\secref}[1]{Section~\ref{#1}}

\begin{document}
\begin{frontmatter}

\title{Platooning of Heterogeneous Vehicles with Actuation Delays: Theoretical and Experimental Results
} 

\author[First]{R. de Haan} 
\author[Second]{L. Redi} 
\author[First]{T. van der Sande}
\author[First]{E. Lefeber}

\address[First]{Mechanical Engineering Department, Eindhoven University of Technology, 
   Eindhoven, The Netherlands (e-mail: r.d.haan@tue.nl, t.p.j.v.d.sande@tue.nl, a.a.j.lefeber@tue.nl).}
\address[Second]{University of Naples Federico II, Naples, Italy (e-mail: l.redi@studenti.unina.it)}

\begin{abstract}                
In this paper we present a prediction-based Cooperative Adaptive Cruise Controller for vehicles with actuation delay, applicable within heterogeneous platoons. We provide a stability analysis for the discrete-time implementation of this controller, which shows the effect of the used sampling times and can be used for selecting appropriate controller gains. The theoretical results are validated by means of experiments using full scale vehicles.
\end{abstract}


\end{frontmatter}

\section{Introduction} \label{sec:introduction}
Cooperative Adaptive Cruise Control (CACC) aims to control the distance of a vehicle with respect to a preceding vehicle. By adopting vehicle-to-vehicle communication in addition to measurements of on-board sensors, CACC can achieve string stable behavior at short inter-vehicle distances \citep{Ploeg_2011}, which means disturbances are attenuated through the vehicle string. Adopting CACC has potential benefits associated with the close following distances. For instance, increased road capacity \citep{Xiao_2018}, reduced fuel consumption \citep{Alam_2010} and increased safety \citep{Li_2017}. However, delays (for instance, actuation delays) have shown to be detrimental to the performance of CACC \citep{Liu_2001, Haan_IFAC_2023} and can compromise the associated benefits.

To deal with delays in platoons, there are approaches that analyze the robustness of control laws to (small) delays \citep{Xiao_2011} or use an approximation of the delay in the controller design \citep{Xing_2016, Haan_ITSC_2023}. To deal with larger delays in platoons, approaches have been presented that compensate the actuation delay. For instance \cite{Davis_2021} compensates the actuation delay for homogeneous platoon and \cite{Bekiaris_2023} uses a prediction based control law to compensate the actuation delay in heterogeneous platoons. Both these approaches have the advantage that they are able to fully compensate the delay, rendering the size of the actuation delay irrelevant. To achieve this full compensation, they either make an assumption on the dynamics of the preceding vehicle (assuming a homogeneous platoon), or they require the vehicle to share information about its driveline. In a practical implementation of CACC, however, it is very likely that the platoon consists of different types of vehicles. Furthermore, communication of the driveline information of the vehicle might pose a problem, since not all required driveline information is currently represented in the existing message definitions from \cite{ensemble_D28} which are commonly accepted as a standard for platooning applications. 

In this paper, we present a controller design where we do not make assumptions on, or require information of the driveline of the preceding vehicle. We use the delay-free controller design from \cite{Lefeber_2020} as a basis for a prediction based controller, to mitigate the effects of the ego vehicle's actuation delay on the error dynamics. The result is a closed-loop system that is input-to-state (ISS) stable with respect to the preceding vehicle's acceleration. As in practice, the preceding vehicle's acceleration is a bounded signal, we can ensure the spacing errors remain within bounds.

Key benefits of the proposed control approach are the fact that no drive-line information of the preceding vehicle is required. The result is a controller that enables the formation of platoons in a heterogeneous setting, since both the drive-line constant, as well as the actuation delay of the preceding vehicle do not affect the response of the vehicle. Since a continuous time prediction based controller is infinite dimensional, we present the discrete-time equivalent by assuming a zero-order hold sampling of the input. The result is a finite dimensional controller, that is validated in experiments with full-scale vehicles.

The paper is outlined as follows. \secref{sec:problem-formulation} introduces the platooning objective and problem formulation. In \secref{sec:controller-design} the continuous time controller design is presented, followed by a stability analysis of the discrete-time implementation of the controller. \secref{sec:experimental-validation} describes the experimental setup and results. Finally, \secref{sec:conclusion} summarizes the conclusions and directions for further research.

\section{Problem formulation}\label{sec:problem-formulation}
\begin{figure}
\begin{center}
\includegraphics[width=8.4cm]{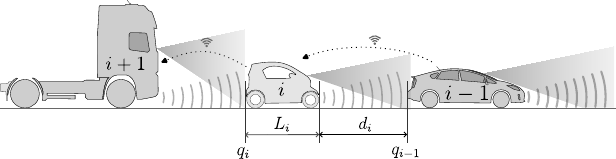}    
\caption{Heterogeneous string of vehicles.} 
\label{fig:heterogeneous-platoon}
\end{center}
\end{figure}

We consider a platoon of $n + 1$ vehicles as depicted by the heterogeneous string of vehicles in \figref{fig:heterogeneous-platoon}. The longitudinal dynamics of each vehicle are modeled according to 
\begin{subequations} \label{eq:vehicle-model}
\begin{align}
\dot{q}_i (t) &= v_i (t) \\
\dot{v}_i (t) &= a_i (t) \\
\dot{a}_i (t) &= -\frac{1}{\tau_i} a_i(t) + \frac{1}{\tau_i} u_i(t - \phi_i), \label{subeq:accleration-model}
\end{align}
\end{subequations}
for $i \in \left\{0, 1, 2, ..., n \right\}$. Here, $q_i$, $v_i$ and $a_i$ are the rear bumper position, velocity and acceleration of vehicle $i$ respectively. The response of vehicle's acceleration $a_i$ subject to input $u_i$ is modeled as a first order system with a time constant $\tau_i >0$ that is associated with the driveline of the vehicle and an actuation delay $\phi_i \geq 0$. Since $\tau_i$ and $\phi_i$ are not necessarily identical for each vehicle, we consider a platoon that is heterogeneous with respect to the driveline.

The control objective of each following vehicle (i.e., $i > 0$) in the platoon is to follow its predecessor at a desired distance $d_{\text{r}, i}$ formulated according to the constant headway spacing policy as
\begin{equation}\label{eq:spacing-policy}
d_{\text{r}, i} = h_i v_i(t) + r_i,
\end{equation}
where $h_i > 0$ is the desired headway to the preceding vehicle, and $r_i \geq 0$ is a constant to account for a distance between the vehicles at standstill. %
By defining the error as the difference between the inter vehicle distance $d_i$ and desired inter vehicle distance $d_{r,i}$ according to \eqref{eq:spacing-policy}, the spacing error $e_i = d_i(t) - d_{r,i}(t) $ is defined as
\begin{equation} \label{eq:spacing-error}
e_i(t) = \left[ q_{i-1}(t) - q_i(t) - L_i\right] - \left[ h_i v_i(t) - r_i  \right]. 
\end{equation}
The goal of each vehicle in the platoon is to control the spacing error \eqref{eq:spacing-error} to zero.

\subsection{Delay-free controller design}
Along the lines of \cite{Lefeber_2020} and \cite{Wijnbergen_2020}, a controller design based on an input-output linearization of the spacing error \eqref{eq:spacing-error} can be followed to obtain a controller for the delay-free plant. 
Defining a coordinate transformation along the lines of \cite{Lefeber_2020} as
\begin{subequations} \label{eq:coordinate-transform}
\begin{align}
x_1(t) &= e_i(t) = q_{i-1}(t) - q_i(t) - h_i v_i(t) - r_i - L_i\notag\\ 
x_2(t) &= \dot{e}_i(t) = v_{i-1}(t) - v_i(t) - h_i a_i(t) \tag{\ref{eq:coordinate-transform}}\\ 
x_3(t) &= v_{i-1}(t) - v_i(t) , \notag 
\end{align}
\end{subequations}
the error \eqref{eq:spacing-error} and its dynamics are given by
\begin{subequations} \label{eq:error-dynamics}
\begin{align}
\dot{x}_1(t) &= x_2(t) \notag\\
\dot{x}_2(t) &= a_{i-1}(t) - \tfrac{h_i - \tau_i}{h_i \tau_i} \left[x_2(t) - x_3(t) \right] -\tfrac{h_i}{\tau_i} u_i(t-\phi_i)\tag{\ref{eq:error-dynamics}}\\
\dot{x}_3(t) &= \tfrac{1}{h_i} x_2(t) - \tfrac{1}{h_i} x_3(t) + a_{i-1}(t). \notag
\end{align}
\end{subequations}%
Consequently, a change of input to the system as
\begin{equation}\label{eq:change-of-input}
u_i(t) = \frac{\tau_i}{h_i} a_{i-1}(t) + (1 - \frac{\tau_i}{h_i}) a_i(t) + \frac{\tau_i}{h_i} \left[ k_p x_1(t) + k_d x_2(t)\right],
\end{equation}
results for the system without delay, i.e., $\phi_i = 0$, in closed-loop dynamics according to
\begin{equation} \label{eq:error-dynamics-delay-free}
\begin{bmatrix}
\dot{x}_1 (t) \\ \dot{x}_2 (t) \\ \dot{x}_3 (t)
\end{bmatrix} =
\begin{bmatrix}
0 & 1 & 0 \\ -k_p & -k_d & 0 \\ 0 & \tfrac{1}{h_i} & -\tfrac{1}{h_i}
\end{bmatrix} \begin{bmatrix}
x_1 (t)\\ x_2 (t)\\ x_3(t)
\end{bmatrix} 
+ \begin{bmatrix}
    0 \\ 0 \\ 1
\end{bmatrix}
a_{i-1}(t).
\end{equation}
For $k_p > 0$, $k_d > 0$ the dynamics \eqref{eq:error-dynamics-delay-free} are ISS with respect to the preceding vehicle's acceleration, i.e., $x(t)$ remains bounded for bounded $a_{i-1}$ and $x(t)$ converges to zero when $a_{i-1}$ is zero.
However, for the plant with actuation delay, i.e., $\phi_i > 0$, the controller \eqref{eq:change-of-input} results in closed-loop dynamics according to
\begin{subequations} \label{eq:error-dynamics-delay}
\begin{equation}
\dot{x}(t) = A_0 x(t)  - A_1 x(t - \phi_i) + \begin{bmatrix}
  0 \\ a_{i-1}(t) - a_{i-1}(t-\phi_i) \\
  a_{i-1}(t)
\end{bmatrix},
\end{equation}
with $x(t) = \begin{bmatrix} x_1 & x_2 & x_3 \end{bmatrix}^\intercal$, and
\begin{equation}
A_0 = \begin{bmatrix}
0 & 1 & 0 \\
0 & \tfrac{\tau_i - h_i}{\tau_i h_i} &  -\tfrac{\tau_i - h_i}{\tau_i h_i} \\
0 & \tfrac{1}{h_i} & - \tfrac{1}{h_i}
\end{bmatrix}, \
A_1 =  \begin{bmatrix}
0 & 0 & 0 \\
- k_p  & \tfrac{h_i - \tau_i}{h_i \tau_i} - k_d & \tfrac{\tau_i - h_i }{h_i \tau_i} \\
0 & 0 & 0
\end{bmatrix}.
\end{equation}
\end{subequations}

In previous work, we analyzed the tuning of the closed-loop system \eqref{eq:error-dynamics-delay} with respect to performance criteria such as string stability, see \cite{Haan_ITSC_2023}. In this work, we pursue a different approach where we aim to compensate the ego vehicle's actuation delay. Using the controller \eqref{eq:change-of-input} as a stabilizing controller for the undelayed system, we use a predictor feedback design approach to compensate the ego vehicle's actuation delay.

\section{Controller design} \label{sec:controller-design}
Predictor feedback design is a method to deal with input delays in (linear) systems. The general idea is to use a stabilizing controller for the undelayed system, and use this controller acting on a prediction of the future states of the system with input delay. A detailed description of general predictor-feedback can be found in \cite{Krstic_2010_delay}.
\subsection{Predictor-feedback based CACC controller}
The predictor-feedback controller based on the delay-free controller \eqref{eq:change-of-input} is given by
\begin{subequations} \label{eq:prediction-controller-continuous}
\begin{align}
 u_i(t) &= \left [ 1 - \tfrac{\tau_i}{h_i} \right] \hat{a}_i^{\phi_i} (t) + \tfrac{\tau_i}{h_i} a_{i-1}(t) - \tfrac{\tau_i}{h_i} \bar{u}_i(t), \label{eq:control-input}\\
 \hat{a}_i^{\phi_i} (t) & = e^{-\tfrac{\phi_i}{\tau_i}} a_i(t) + \int_{t - \phi_i}^{t} \tfrac{1}{\tau_i} e^{\left(- \tfrac{t-s}{\tau_i} \right)}  u_i(s) \mathrm{d} s \\
\bar{u}_i (t) &= - \begin{bmatrix}
k_p & k_d
\end{bmatrix} \hat{x}^{\phi_i} (t) \\
\hat{x}^{\phi_i} (t) & = \begin{bmatrix}
1 & \phi_i \\ 0 & 1
\end{bmatrix} \begin{bmatrix}
x_1(t) \\ x_2(t)
\end{bmatrix} + \int_{t - \phi_i}^{t} \begin{bmatrix}
t-s \\ 1
\end{bmatrix} \bar{u}_i(s) \mathrm{d} s . \label{eq:prediction-feedback}
\end{align}
\end{subequations}

Applying the controller~\eqref{eq:prediction-controller-continuous} on the system~\eqref{eq:error-dynamics}, results for $t> \phi_i$ in closed-loop dynamics 
\begin{equation} \label{eq:closed-loop-prediction-continuous}
\begin{bmatrix}
\dot{x}_1(t) \\
\dot{x}_2(t) \\
\dot{x}_3(t)
\end{bmatrix} = 
\begin{bmatrix}
0 & 1 & 0 \\
-k_p & -k_d & 0 \\
0 & \tfrac{1}{h_i} & -\tfrac{1}{h_i}
\end{bmatrix}
\begin{bmatrix}
x_1(t) \\ 
x_2(t) \\
x_3(t)
\end{bmatrix} 
+
\begin{bmatrix}
0 & 0\\
1 & - 1 \\
1 & 0
\end{bmatrix}
\begin{bmatrix}
a_{i-1} (t) \\ a_{i-1}(t-\phi_i) 
\end{bmatrix},
\end{equation}
which is input-to-state-stable (ISS) with respect to the predecessors acceleration $a_{i-1}(t)$ for $k_p > 0$, $k_d > 0$. That is, $x(t)$ remains bounded for bounded $a_{i-1}(t)$, and $x(t)$ converges to zero when $a_{i-1}(t)$ converges to zero. Due to the term $a_{i-1}(t) - a_{i-1}(t - \phi_i)$ that acts as disturbance to the error dynamics, there will be an error induced when the leader vehicle accelerates, which was not present in the delay-free case. By using a prediction for $a_{i-1}(t+\phi_i)$ instead of $a_{i-1}(t)$ in \eqref{eq:control-input}, this error can be mitigated. However, such a prediction is considered outside the scope of this paper, as we focus on compensating the effect of the ego-vehicle's driveline on the stabilization of the error dynamics.

\subsection{Discrete time controller}\label{sec:discrete-controller}
In practice, a controller is typically executed on a computer platform with the control action being applied to the vehicle at discrete times.
Using a sampling time $T_s$, assuming zero-order-hold (ZOH) for the input $u_i$ such that
\begin{equation} \label{eq:zoh}
u_i(t) = u_i(kTs) \text{ for } kT_s \leq t < kT_s + T_s, \ k\in \mathbb{N},    
\end{equation}
and assuming the delay $\phi_i = d T_s$ for some integer $d$, the discrete-time equivalent of the controller \eqref{eq:prediction-controller-continuous} is given by
\begin{subequations} \label{eq:prediction-controller-discrete}
\begin{equation}
    u_i(k T_s) = \left[ 1 - \tfrac{\tau_i}{h_i} \right] \hat{a}_i^{\phi_i} (kT_s) +  \tfrac{\tau_i}{h_i} a_{i-1}(kT_s) -  \tfrac{\tau_i}{h_i} \bar{u}_i (k T_s) 
\end{equation}
where
\begin{align}
    \hat{a}_i^{\phi_i}(k T_s)&= e^{- \tfrac{dT_s}{\tau_i}} a_i(k T_s) \notag\\
    & + \sum_{j=1}^{d} \left( e^{\tfrac{-(j-1)T_s}{\tau_i}} - e^{\tfrac{-jT_s}{\tau_i}} \right) u_i(kT_s - jT_s) \\
    \bar{u}_i(kT_s) &= - \left[ k_p \ k_d \right] \hat{x}^{\phi_i}(k T_s) \\
    \hat{x}^{\phi_i}(k T_s) &= \begin{bmatrix} 1 & dT_s \\ 0 & 1\end{bmatrix} \begin{bmatrix}
        x_1(kT_s) \\ x_2(kT_s)
    \end{bmatrix} \notag \\ & + \sum_{j=1}^{d} \begin{bmatrix}
        \tfrac{1}{2}T_s^2 + (j -1) T_s^2 \\
        T_s
    \end{bmatrix}
    \bar{u}_i(kT_s - jT_s).
    \end{align}
\end{subequations}

Using the discrete time time representation \eqref{eq:prediction-controller-discrete} has some advantages. For instance, the discrete-time controller is finite dimensional as only a finite number of past inputs has to be remembered to compute the control action. Contrary to the continuous time controller \eqref{eq:prediction-controller-continuous} which is infinite-dimensional, since it contains the distributed delay term involving past controls.

\subsection{Closed-loop stability}
For analysis of the closed-loop stability of vehicle $i$ employing controller \eqref{eq:prediction-controller-discrete}, we consider the continuous time error dynamics 
\begin{subequations} \label{eq:error-dynamics-with-disturbance}
\begin{align}
\dot{x}(t) = A_0 x(t) + B_1 u_i(t-\phi_i) + B_2 a_{i-1} (t),
\end{align}
where $x(t)$ and $A_0$ are defined as in \eqref{eq:error-dynamics-delay}, and 
\begin{align}
B_1 = \begin{bmatrix}
    0 & - \tfrac{h_i}{\tau_i} & 0
\end{bmatrix}^\intercal , &\  B_2 = \begin{bmatrix}
    0 & 1 & 1
\end{bmatrix}^\intercal.
\end{align}
\end{subequations}
The input $u_i(t-\phi_i)$ is a discrete signal with ZOH according to \eqref{eq:zoh}. However, we consider the acceleration $a_{i-1}(t)$ as a continuous exogenous disturbance. The evolution of the dynamics~\eqref{eq:error-dynamics-with-disturbance} in discrete-time can subsequently be described by 
\begin{subequations} \label{eq:discrete-error-dynamics-with-disturbance-input}
\begin{align}
    x(kT_s + T_s) = \Phi x(kT_s) + \Gamma u_i(kT_s - dT_s) + w(kT_s),
\end{align}
where
\begin{align}
\Phi &= e^{AT_s}, \quad  \Gamma = \int_{0}^{T_s} e^{A_0s}\mathrm{d}s B_1, \\ w(kT_s) &= \int_{0}^{T_s} e^{A_0s} B_2 a_{i-1}(kT_s + T_s - s) \mathrm{d}s.
\end{align}
\end{subequations}
By introducing $\bar{x}(kT_s)$ and $z(kT_s)$ as
\begin{align}
\bar{x}(kT_s) = \begin{bmatrix}
    x(kT_s) \\
    u_i(kT_s - dT_s) \\
    \vdots \\
    u_i(kT_s - T_s)
\end{bmatrix}, \ & z(kT_s) = \begin{bmatrix}
    \bar{u}_i(kT_s - dT_s) \\
    \vdots \\
    \bar{u}_i(kT_s - T_s)
\end{bmatrix},
\end{align}
the dynamics \eqref{eq:discrete-error-dynamics-with-disturbance-input} can be written as 
\begin{subequations} \label{eq:error-dynamics-discrete-with-disturbance}
\begin{equation}
    \bar{x}(kT_s+T_s) = \bar{A} \bar{x}(kT_s) + \bar{B}_1 u_i(kT_s) + \bar{B}_2 w(kT_s),
\end{equation}
where 
\begin{align}
    \bar{A} = \begin{bmatrix}
        \Phi & \Gamma & 0 & 0 & \cdots & 0 \\
        0 & 0 & 1 & 0 & \cdots & 0 \\
        \vdots & \ddots & \ddots & \ddots & \ddots & \vdots \\
        0 & \cdots & 0 & 0 & 1 & 0 \\
        0 & \cdots & 0 & 0 & 0 & 1 \\
        0 & \cdots & 0 & 0 & 0 & 0 \\
    \end{bmatrix}, &
    \bar{B}_1 = \begin{bmatrix}
        0 \\ 0 \\ \vdots \\ 0 \\ 0 \\ 1
    \end{bmatrix}, & \bar{B}_2 = \begin{bmatrix}
        I \\ 0 \\ \vdots \\ 0 \\ 0  \\ 0
    \end{bmatrix}.
\end{align}
\end{subequations}
Similarly, the controller \eqref{eq:prediction-controller-discrete} can be written in the form 
\begin{subequations} \label{eq:controller-linear-form}
\begin{align}
u_i(kT_s) &= C_x \bar{x}(kT_s) + C_z z(kT_s) + C_a a_{i-1}(kT_s) \\
z(kT_s + T_s) &=  A_x \bar{x}(k T_s) + A_z z(kT_s) + B_a a_{i-1}(kT_s).
\end{align}
\end{subequations}


Combining $\eqref{eq:controller-linear-form}$ and \eqref{eq:error-dynamics-discrete-with-disturbance}, the closed-loop dynamics of vehicle $i$ employing the controller \eqref{eq:prediction-controller-discrete} can be given in the form 
\begin{subequations}
\begin{equation}
 x_{cl}(kT_s + T_s) = A_{cl} x_{cl}(kT_s) + B_{cl}  w_{cl}(kT_s),
\end{equation}
where
\begin{align}
    x_{cl}(kT_s) = \begin{bmatrix}
        \bar{x}(kT_s) \\ z(kT_s)
    \end{bmatrix}, \ & w_{cl}(kT_s) = \begin{bmatrix}
        w(kT_s) \\
        a_{i-1}(kT_s)
    \end{bmatrix}.
\end{align}
\end{subequations}
Note that bounded $a_{i-1}(t)$ implies bounded $w_{cl}(kT_s)$, and for $a_{i-1}(t)$ converging to zero, $w_{cl}(kT_s)$ goes to zero. Finally, for the system to be ISS with respect to the disturbance~$w_{cl}$, all eigenvalues of $A_{cl}$ should be within the unit-disk. 

\section{Experimental validation} \label{sec:experimental-validation}
\begin{figure}
\begin{center}
\includegraphics[width=8.4cm]{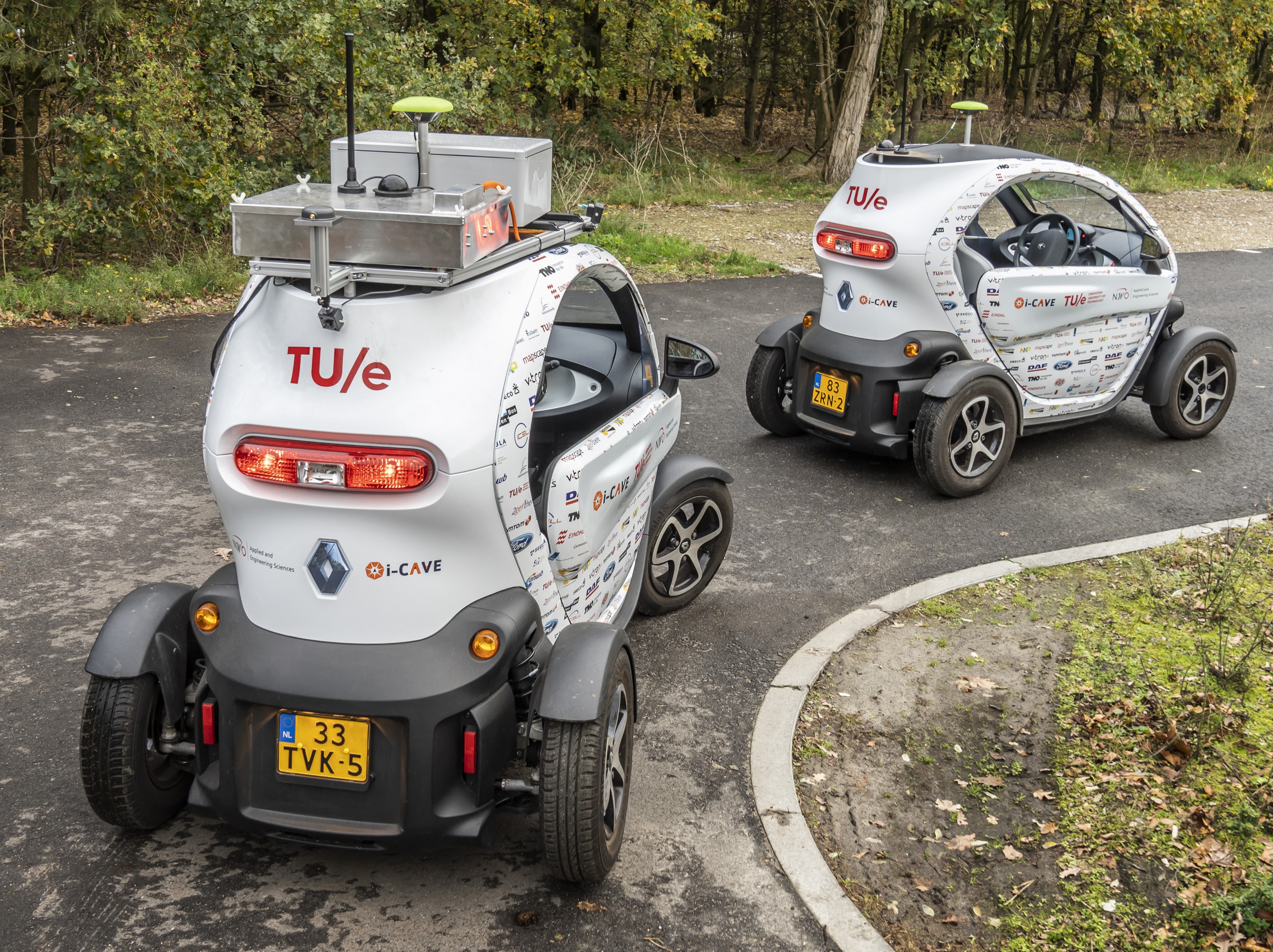}    
\caption{Platoon consisting of two experimental vehicles.} 
\label{fig:experimental-vehicles}
\end{center}
\end{figure}

\begin{figure}
\begin{center}
\includegraphics[width=8.4cm]{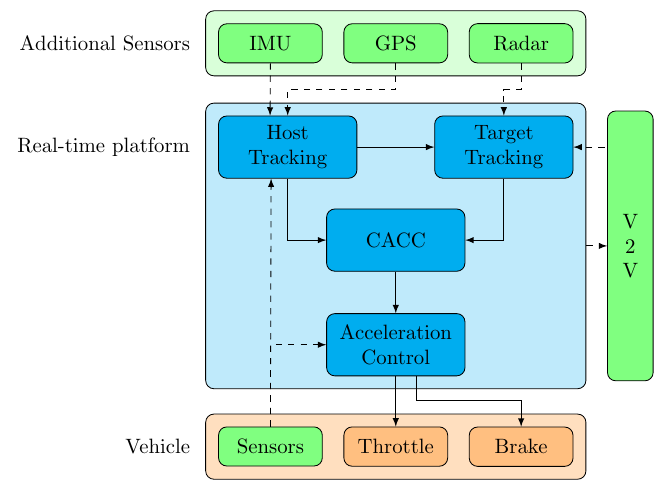}    
\caption{Schematic overview of the automated vehicle and its soft- and hardware components.} 
\label{fig:schematic-overview-twizy}
\end{center}
\end{figure}

To evaluate the proposed controller, experiments are performed with a platoon consisting of the two full-scale vehicles shown in \figref{fig:experimental-vehicles}. A schematic overview of the components of the longitudinal automation of the vehicle is depicted in \figref{fig:schematic-overview-twizy}. The CACC controller obtains the measurements from the sensors as listed in \tabref{tab:measurement-sample-times}. To reconstruct the error $e_i = x_1$, measurements of the radar and speedometer are used. The error derivative $x_2$, is based on measurements of the radar and IMU sensor. The preceding vehicle's acceleration is measured on-board of the leader vehicle with an IMU and obtained on the ego vehicle by means of V2V communication. Therefore, V2V is considered a sensor in \tabref{tab:measurement-sample-times}. The controller~\eqref{eq:prediction-controller-discrete} is executed with sample time $T_s = 10$ ms and provides the acceleration setpoint $u_i(t)$ at a frequency of 100 Hz to the vehicle. A more detailed description of the automation of the vehicles can be found in \cite{Hoogeboom_2020}. 

\renewcommand{\arraystretch}{1.5}
\begin{table}[htb]
\centering
\caption{Measurements performed by associated sensors on the ego vehicle.}
\label{tab:measurement-sample-times}
\begin{tabular}{| l | l l |}\hline 
Sensor & \multicolumn{2}{l| }{Measurement} \\ \hline \hline
\multirow{2}{*}{Radar} & Range & $d_i$ \\
 & Range rate & $\dot{d}_i$ \\ \hline
Speedometer & Ego velocity & $v_i$ \\ \hline
IMU & Ego acceleration & $a_i$ \\ \hline
V2V & Leader acceleration & $a_{i-1}$\\ \hline
\end{tabular}
\end{table}

\subsection{Experiment design}
To validate the proposed controller, the closed-loop response of the experimental vehicle employing the prediction controller \eqref{eq:prediction-controller-discrete} is considered. The two vehicles are initialized at a certain initial inter vehicle distance in standstill. The standstill distance $r_i$ is chosen such that the error \eqref{eq:spacing-error} is zero when activating the controller \eqref{eq:prediction-controller-discrete}. When the controller has been active for a time $t_\text{active} > \phi_i$, an error is introduced by adjusting the standstill distance $r_i$ at time $t_0$. The resulting response to an initial condition of $x(t_0)~=~\begin{bmatrix}
e_0 & 0 & 0
\end{bmatrix}^\intercal$ of the experimental vehicle is compared to a simulation of the systems \eqref{eq:closed-loop-prediction-continuous} and \eqref{eq:error-dynamics-delay} subject to identical initial conditions. Both in the experiment and simulation, a time constant $\tau_i = 0.067 \text{ seconds}$ and actuation delay $\phi_i = 0.15 \text{ seconds}$ delay of the experimental vehicle are used in combination with a headway $h_i = 0.5 \text{ seconds}$. The controller gains are set at $k_p = 0.2$ and $k_d = 0.7 - k_p \tau_i$. However, this particular choice of gains is arbitrary (and chosen to be identical to the gains in \cite{Haan_ITSC_2023}). Detailed controller tuning, taking into account passenger comfort and string stability is planned for future research. The main purpose of these experiments is to validate the closed-loop response of the experimental vehicle is similar to the theoretical response, to confirm the assumptions and model are correct.

\subsection{Experimental results}
\begin{figure}
\begin{center}
\begin{subfigure}{\linewidth}
\input{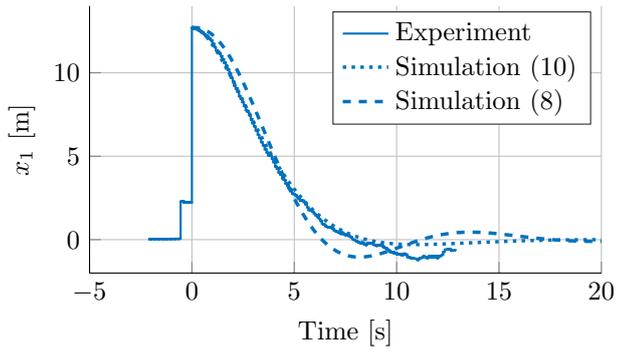}
\caption{Error $x_1$.}
\end{subfigure}
\begin{subfigure}{\linewidth}
\vspace{2mm}
\input{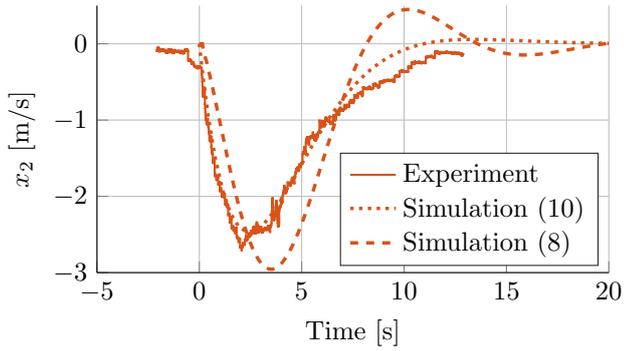}
\caption{Error derivative $x_2$.}
\end{subfigure}
\begin{subfigure}{\linewidth}
\vspace{2mm}
\definecolor{mycolor1}{rgb}{0.92900,0.69400,0.12500}%
\begin{tikzpicture}

\begin{axis}[%
width=0.761\linewidth,
height=0.4\linewidth,
at={(0\linewidth,0\linewidth)},
scale only axis,
xmin=-5,
xmax=20,
xlabel style={font=\color{white!15!black}},
xlabel={Time [s]},
ymin=-3,
ymax=0.5,
ylabel style={font=\color{white!15!black}},
ylabel={$x_3$ [m/s]},
axis background/.style={fill=white},
axis x line*=bottom,
axis y line*=left,
xmajorgrids,
ymajorgrids,
legend style={legend cell align=left, align=left, draw=white!15!black},
ylabel near ticks,
xlabel near ticks,
scaled ticks=false, tick label style={/pgf/number format/fixed},
legend style={at={(0.98,0.025)}, anchor = south east, legend cell align=left, align=left, draw=white!15!black}
]
\addplot[const plot, color=mycolor1, line width=0.8pt] table[row sep=crcr] {%
-2.1	-0.0150760756805539\\
-2.08000000000002	-0.0100351460278034\\
-2.03999999999994	-0.010235314257443\\
-1.99999999999998	-0.00757037615403533\\
-1.96000000000001	-0.00679037300869823\\
-1.91999999999994	-0.00801830645650625\\
-1.87999999999997	-0.00468642311170697\\
-1.84000000000001	-0.00533545389771461\\
-1.79999999999993	-0.00653604371473193\\
-1.75999999999997	-0.0070476233959198\\
-1.72	-0.0074714059010148\\
-1.68000000000004	-0.00517796911299229\\
-1.63999999999996	-0.0066933361813426\\
-1.6	-0.00785253662616014\\
-1.56000000000004	-0.0100379940122366\\
-1.51999999999996	-0.0104176998138428\\
-1.48	-0.00621293904259801\\
-1.44000000000003	-0.00767565006390214\\
-1.39999999999995	-0.00945010967552662\\
-1.35999999999999	-0.00944204162806273\\
-1.32000000000003	-0.0109996944665909\\
-1.27999999999995	-0.0130145726725459\\
-1.23999999999999	-0.0100092208012938\\
-1.20000000000002	-0.00958448369055986\\
-1.15999999999995	-0.00999591592699289\\
-1.11999999999998	-0.0111457444727421\\
-1.08000000000002	-0.00782578717917204\\
-1.03999999999994	-0.00940718036144972\\
-0.999999999999977	-0.0106165232136846\\
-0.960000000000013	-0.0118101537227631\\
-0.919999999999936	-0.0138403428718448\\
-0.879999999999972	-0.00950959324836731\\
-0.840000000000009	-0.0105497455224395\\
-0.799999999999931	-0.0120361335575581\\
-0.759999999999968	-0.0148358773440123\\
-0.720000000000004	-0.0154753476381302\\
-0.680000000000041	-0.00910121574997902\\
-0.639999999999963	-0.00921955704689026\\
-0.6	-0.00858594570308924\\
-0.560000000000036	-0.00913797039538622\\
-0.549999999999931	-0.125\\
0.110000000000037	-0.1875\\
0.270000000000005	-0.3125\\
0.350000000000046	-0.375\\
0.419999999999982	-0.4375\\
0.490000000000032	-0.5\\
0.580000000000064	-0.625\\
0.659999999999991	-0.75\\
0.750000000000023	-0.9375\\
0.830000000000064	-1\\
0.919999999999982	-1.0625\\
1.00000000000002	-1.1875\\
1.09000000000005	-1.25\\
1.16999999999998	-1.3125\\
1.25000000000002	-1.4375\\
1.33000000000006	-1.5\\
1.40999999999999	-1.625\\
1.48000000000004	-1.75\\
1.56999999999996	-1.875\\
1.65999999999999	-1.9375\\
1.74000000000003	-2\\
1.81999999999996	-2.125\\
1.90999999999999	-2.1875\\
1.99000000000003	-2.3125\\
2.05999999999997	-2.375\\
2.22000000000005	-2.3125\\
2.29999999999998	-2.375\\
2.37000000000003	-2.3125\\
2.45000000000007	-2.375\\
2.53	-2.3125\\
2.87000000000003	-2.375\\
3.03	-2.4375\\
3.46000000000006	-2.375\\
3.54999999999998	-2.3125\\
3.64000000000001	-2.25\\
3.72000000000005	-2.3125\\
3.80999999999997	-2.4375\\
3.9	-2.3125\\
3.98000000000004	-2.1875\\
4.06999999999996	-2.125\\
4.14000000000001	-2.0625\\
4.30999999999997	-2\\
4.65	-1.9375\\
4.79999999999998	-1.875\\
4.96000000000006	-1.8125\\
5.03999999999999	-1.6875\\
5.12000000000003	-1.625\\
5.20000000000007	-1.5\\
5.37000000000003	-1.4375\\
5.53999999999999	-1.375\\
5.78	-1.3125\\
5.87000000000003	-1.1875\\
6.11000000000004	-1.125\\
6.52	-1.0625\\
6.67999999999997	-1\\
6.85000000000005	-0.9375\\
7.09000000000005	-0.875\\
7.33000000000006	-0.8125\\
7.56999999999996	-0.75\\
7.74000000000003	-0.6875\\
7.99000000000003	-0.625\\
8.49000000000003	-0.5625\\
8.80999999999997	-0.5\\
9.53999999999999	-0.4375\\
9.78999999999999	-0.375\\
10.2000000000001	-0.3125\\
10.36	-0.25\\
11.06	-0.1875\\
11.51	-0.125\\
12.89	-0.125\\
};
\addlegendentry{Experiment}

\addplot [color=mycolor1, dotted, line width=1.3pt]
  table[row sep=crcr]{%
0	0\\
0.0123321568313308	-0.000382350723004521\\
0.0617398364440263	-0.00917082676243908\\
0.206490445023697	-0.0903493574177183\\
0.436333613784576	-0.331582793104307\\
1.14613182029	-1.29610164523542\\
1.54613182029	-1.74774142327866\\
1.94613182029	-2.07548647576215\\
2.34613182029	-2.28158652542472\\
2.74613182029	-2.3812527913756\\
3.14613182029	-2.39355622062818\\
3.54613182029	-2.33746742894397\\
3.94613182029	-2.23022736796113\\
4.34613182029	-2.08677638790439\\
4.74613182029	-1.91966274559379\\
5.14613182029	-1.73916484370837\\
5.94613182029	-1.36908671527328\\
6.34613182029	-1.19075567778733\\
6.74613182029	-1.02202296084033\\
7.14613182029001	-0.865291134379209\\
7.54613182029	-0.722051553235271\\
7.94613182029001	-0.593061191566676\\
8.34613182029	-0.478497712514891\\
8.74613182029001	-0.378093557813475\\
9.14613182029001	-0.291250302209125\\
9.54613182029001	-0.217134797705608\\
9.94613182029001	-0.154758787344427\\
10.34613182029	-0.10304373189491\\
10.74613182029	-0.0608725893911952\\
11.14613182029	-0.0271302348693681\\
11.54613182029	-0.000734119975227543\\
11.94613182029	0.01934333944088\\
12.34613182029	0.0340592860477926\\
12.74613182029	0.0442906265020468\\
13.14613182029	0.0508288282824694\\
13.54613182029	0.0543783504689515\\
13.94613182029	0.0555578795541329\\
14.34613182029	0.0549036621603243\\
15.14613182029	0.0498567784114741\\
16.34613182029	0.0378037696893507\\
17.94613182029	0.0215174010749202\\
19.14613182029	0.0121398142426514\\
20	0.00727234091030127\\
};
\addlegendentry{Simulation \eqref{eq:closed-loop-prediction-continuous}}

\addplot [color=mycolor1, dashed, line width=1.3pt]
  table[row sep=crcr]{%
0	0\\
0.160149823967185	-0.000258741535301255\\
0.190059415355112	-0.00340529935077427\\
0.228142511917582	-0.010951908250167\\
0.280264184176161	-0.0248743156901838\\
0.344852738927447	-0.0453628431255844\\
0.411298170753927	-0.0729381544437118\\
0.502281382015568	-0.120474449490597\\
0.593264593277208	-0.17702224792103\\
0.701332901326335	-0.255537951767149\\
0.814896209871105	-0.349374219119287\\
0.951750706343674	-0.475511093883032\\
1.0882277814612	-0.612762489390136\\
1.22470485657872	-0.758803750790761\\
1.36048620157035	-0.910344435781433\\
1.62651023361003	-1.21667113964837\\
1.75675292065809	-1.3674523413991\\
1.91539411773454	-1.54872911775492\\
2.03310514868334	-1.6801030300249\\
2.15081617963214	-1.80771597723639\\
2.29620243027858	-1.95886195707197\\
2.44158868092502	-2.10153239906949\\
2.59837057483615	-2.24441588692459\\
2.75335299383363	-2.37319850961445\\
2.89596581514111	-2.47982945732434\\
3.04517148414566	-2.57851887510835\\
3.231056450093	-2.68231652680492\\
3.41000548456826	-2.76155007730029\\
3.56624949521187	-2.81382199467564\\
3.75053769553227	-2.85530059254678\\
3.90591485030841	-2.87356871425057\\
4.06129200508455	-2.87692912803678\\
4.22668250860995	-2.86462175171151\\
4.39035255984972	-2.83697793085318\\
4.53584524939995	-2.80017737307621\\
4.67702761524833	-2.75406609910457\\
4.84837639987189	-2.68517913326692\\
4.98191330038552	-2.62240275877428\\
5.11545020089916	-2.55233039443861\\
5.28897691243496	-2.45131339026021\\
5.41971544312614	-2.3685393883582\\
5.55045397381732	-2.28074351428295\\
5.73557080517325	-2.14892693512848\\
5.86020413138359	-2.05599628619371\\
5.98483745759393	-1.96034275061271\\
6.1418265853408	-1.83670212232103\\
6.42794129280331	-1.60536180353709\\
6.86933858681691	-1.24474604642456\\
7.09010386970164	-1.06764272460736\\
7.20833680519027	-0.974751295434775\\
7.32656974067891	-0.883601996919747\\
7.52435373369548	-0.73562346614812\\
7.66070609919012	-0.637370916633561\\
7.79705846468476	-0.542624291811816\\
7.99146241138249	-0.414073529060413\\
8.13992364704867	-0.321465874106195\\
8.28838488271484	-0.234010723473425\\
8.4795151604808	-0.129279992584493\\
8.69470831094451	-0.0222658684792805\\
8.86803554602447	0.0551614375514156\\
9.01854992849992	0.115984527609438\\
9.24729751697086	0.197216383459715\\
9.41716330946537	0.248946474340467\\
9.58702910195989	0.293518916353616\\
9.76613619441228	0.332950774948795\\
9.94933924189347	0.36551453895256\\
10.1163201414759	0.388684459673076\\
10.2716057728202	0.40493757821427\\
10.4387446036249	0.416989372056154\\
10.6238747787339	0.4241271631426\\
10.7926220070512	0.425405939192668\\
10.9405263929845	0.422794714655797\\
11.0972624732717	0.416483756643789\\
11.2916748795271	0.404001209764736\\
11.4700443220586	0.388561309090896\\
11.6129138243964	0.373899047019652\\
11.7564231026934	0.357311254152702\\
11.9647827140927	0.330287733211868\\
12.2482614806045	0.289298816374846\\
12.6035200379687	0.233355700868813\\
13.4458502724143	0.0972067841440847\\
13.8251369662198	0.040799174441684\\
13.973791178457	0.0203945623072563\\
14.1533746726695	-0.00287770495744866\\
14.4027246737895	-0.0325735001152765\\
14.6934065758557	-0.062577439938476\\
14.937596637542	-0.084009477837359\\
15.156005211185	-0.100234292432862\\
15.374413784828	-0.113616230118932\\
15.5847463492013	-0.123801239040539\\
15.8332817574633	-0.132527242081551\\
16.1278007270048	-0.138935299524288\\
16.3636545941833	-0.141013491858903\\
16.7108845891598	-0.139608507007519\\
16.898389344109	-0.13691081884674\\
17.3011387140112	-0.127161063799818\\
17.6354300098777	-0.116006326340926\\
18.1916868571832	-0.0928818761521306\\
19.7143479013661	-0.0238734281601438\\
20	-0.0127248671307498\\
};
\addlegendentry{Simulation \eqref{eq:error-dynamics-delay}}

\end{axis}
\end{tikzpicture}%
\caption{Internal dynamic state $x_3$.}
\end{subfigure} 
\begin{subfigure}{\linewidth}
\vspace{2mm}
\input{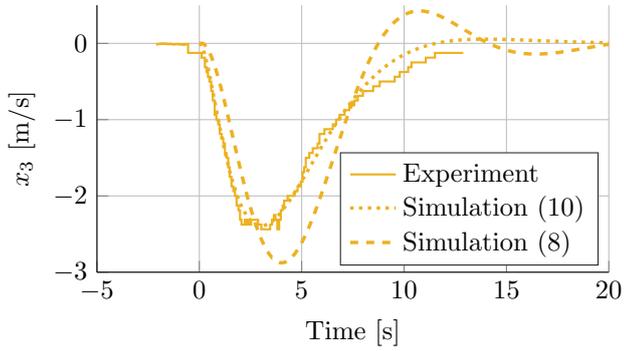}
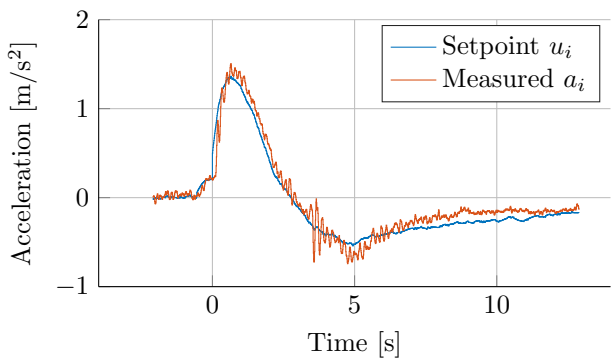
\caption{Acceleration $a_i$ subject to input $u_i$.}
\label{fig:acceleration-experiments}
\end{subfigure}
\caption{Measured experimental response of vehicle deploying predictor feedback based controller \eqref{eq:prediction-controller-discrete}, compared to simulated response of \eqref{eq:error-dynamics-delay} and \eqref{eq:closed-loop-prediction-continuous}.} 
\label{fig:prediction-based-controller-experimental-response}
\end{center}
\end{figure}

The measured experimental response of the experimental vehicle employing the discrete-time prediction based controller \eqref{eq:prediction-controller-discrete} is depicted in \figref{fig:prediction-based-controller-experimental-response}. The response subject to the initial condition in the error at $t_0 = 0$, nicely resembles the theoretical closed-loop response of \eqref{eq:closed-loop-prediction-continuous}. Especially on the interval $t \in [0, 7]$ seconds, all three states match the simulated response. For $t>7$ seconds, there is a mismatch present between the experimental response and simulation. This mismatch can be explained by the fact that at low velocities ($v_i < 0.5$ m/s) the requested deceleration was not achieved by the vehicle, as shown in \figref{fig:acceleration-experiments}. The vehicle was configured to perform all braking by the electric motor to have the response according to the assumed model \eqref{eq:vehicle-model} for both positive and negative accelerations. For future experiments the braking using the mechanical disc-brakes could be used in addition to the regenerative braking of the motor, to make sure the vehicle achieves the desired negative acceleration setpoint at all times. However, this might result in different characteristics (actuation delay and time constant) for the vehicle to a negative acceleration setpoint compared to a positive setpoint, resulting in a more complex vehicle model.

\begin{figure}
\begin{center}
\begin{subfigure}{\linewidth}
\input{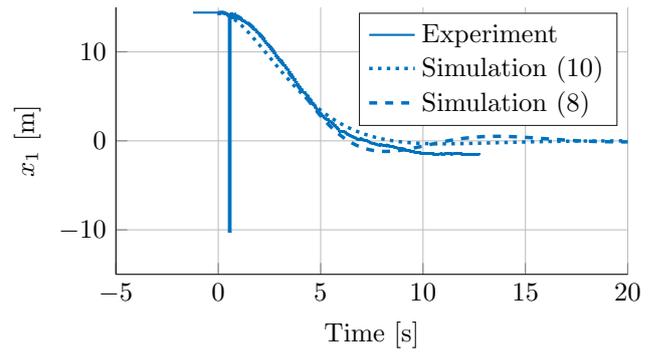}
\caption{Error $x_1$.}
\label{fig:conventional-controller-experimental-response-error}
\end{subfigure}\vspace{2mm}
\begin{subfigure}{\linewidth}
\input{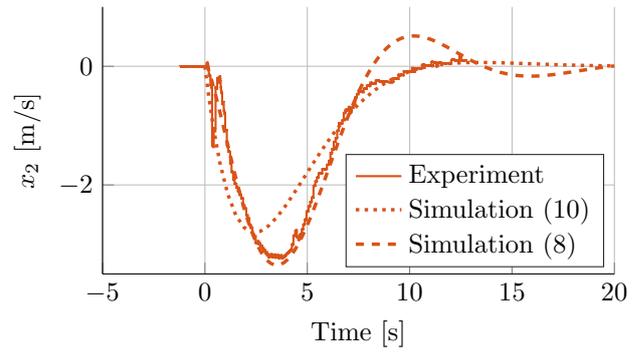}
\caption{Error derivative $x_2$.}
\end{subfigure}\vspace{2mm}
\begin{subfigure}{\linewidth}
\definecolor{mycolor1}{rgb}{0.92900,0.69400,0.12500}%
\begin{tikzpicture}

\begin{axis}[%
width=0.761\linewidth,
height=0.4\linewidth,
at={(0\linewidth,0\linewidth)},
scale only axis,
xmin=-5,
xmax=20,
xlabel style={font=\color{white!15!black}},
xlabel={Time [s]},
ymin=-3.5,
ymax=0.5,
ylabel style={font=\color{white!15!black}},
ylabel={$x_3$ [m/s]},
axis background/.style={fill=white},
axis x line*=bottom,
axis y line*=left,
xmajorgrids,
ymajorgrids,
legend style={legend cell align=left, align=left, draw=white!15!black},
ylabel near ticks,
xlabel near ticks,
scaled ticks=false, tick label style={/pgf/number format/fixed},
legend style={at={(0.98,0.025)}, anchor = south east, legend cell align=left, align=left, draw=white!15!black}
]
\addplot[const plot, color=mycolor1, line width=0.8pt] table[row sep=crcr] {%
-1.23	0\\
0.090000000000007	0.0625\\
0.190000000000001	0\\
0.27	-0.125\\
0.349999999999998	-1.0625\\
0.429999999999996	-0.75\\
0.510000000000009	-0.699431598186493\\
0.52	0\\
0.590000000000007	0\\
0.599999999999998	0.1875\\
0.679999999999996	0.0625\\
0.77	0\\
0.840000000000007	-0.1875\\
0.920000000000005	-0.3125\\
0.989999999999998	-0.5625\\
1.07	-0.8125\\
1.14	-0.875\\
1.23000000000001	-1.0625\\
1.3	-1.1875\\
1.38	-1.3125\\
1.45000000000001	-1.375\\
1.54	-1.4375\\
1.61	-1.5\\
1.69	-1.5625\\
1.77	-1.625\\
1.86	-1.75\\
1.94	-1.8125\\
2.02	-1.9375\\
2.09000000000001	-2\\
2.17000000000001	-2.125\\
2.24	-2.1875\\
2.32	-2.3125\\
2.39	-2.375\\
2.48000000000001	-2.5\\
2.63	-2.5625\\
2.71	-2.625\\
2.8	-2.6875\\
2.87000000000001	-2.75\\
2.95000000000001	-2.8125\\
3.03	-2.875\\
3.11	-2.9375\\
3.28	-3\\
3.44	-3.0625\\
3.6	-3.125\\
3.76000000000001	-3.1875\\
4.25	-3.125\\
4.32	-3\\
4.41	-2.9375\\
4.49	-3.0625\\
4.56000000000001	-3.125\\
4.64	-3.0625\\
4.73000000000001	-3\\
4.8	-2.9375\\
4.88	-2.875\\
4.96	-2.75\\
5.06000000000001	-2.6875\\
5.14	-2.625\\
5.22	-2.5625\\
5.3	-2.375\\
5.39	-2.3125\\
5.47	-2.25\\
5.55	-2.1875\\
5.73000000000001	-2.125\\
5.96	-2.0625\\
6.05	-2\\
6.13	-1.875\\
6.21	-1.8125\\
6.29	-1.6875\\
6.38	-1.625\\
6.46	-1.5\\
6.54	-1.375\\
6.71	-1.3125\\
6.78	-1.1875\\
6.86	-1.125\\
7.02	-1.0625\\
7.09000000000001	-1.125\\
7.17000000000001	-1.0625\\
7.25	-1\\
7.34000000000001	-0.875\\
7.42000000000001	-0.8125\\
7.5	-0.75\\
7.75	-0.6875\\
7.91	-0.625\\
8.08	-0.5625\\
8.34000000000001	-0.5\\
9.21	-0.4375\\
9.46	-0.375\\
9.71	-0.3125\\
10.11	-0.25\\
10.19	-0.3125\\
10.27	-0.25\\
10.6	-0.1875\\
11.07	-0.125\\
11.3	-0.1875\\
11.38	-0.125\\
12.04	-0.0625\\
12.43	0\\
12.52	-0.0625\\
12.76	-0.0625\\
};
\addlegendentry{Experiment}

\addplot [color=mycolor1, dotted, line width=1.3pt]
  table[row sep=crcr]{%
0	0\\
0.0108772875313115	-0.000337682306035703\\
0.0544561638586814	-0.00814137158180372\\
0.192343152879719	-0.0899775824656643\\
0.415888263642852	-0.347446628344734\\
0.721364315716482	-0.812716296416912\\
1.11246519129434	-1.42062500043951\\
1.51246519129434	-1.943674673459\\
1.91246519129434	-2.32730370282832\\
2.31246519129434	-2.57202656569001\\
2.71246519129434	-2.69436722186757\\
3.11246519129434	-2.71580526858376\\
3.51246519129434	-2.65795358340134\\
3.91246519129434	-2.54056009682289\\
4.31246519129434	-2.380791940607\\
4.71246519129434	-2.19310081493131\\
5.11246519129434	-1.98934805611971\\
5.91246519129434	-1.56960367085539\\
6.31246519129434	-1.36666684150932\\
6.71246519129434	-1.17432877557139\\
7.11246519129434	-0.995410281551536\\
7.51246519129434	-0.831681065288095\\
7.91246519129434	-0.684062310549624\\
8.31246519129434	-0.552804541450204\\
8.71246519129434	-0.437641600233537\\
9.11246519129434	-0.337922116854706\\
9.51246519129434	-0.252720178759439\\
9.91246519129434	-0.180927094902195\\
10.3124651912943	-0.121326227737786\\
10.7124651912943	-0.0726528685707706\\
11.1124651912943	-0.0336410760538719\\
11.5124651912943	-0.00305930103577978\\
11.9124651912943	0.0202635036633687\\
12.3124651912943	0.0374197322230749\\
12.7124651912943	0.0494115288073864\\
13.1124651912943	0.0571444942523343\\
13.5124651912943	0.0614255983880234\\
13.9124651912943	0.0629643462597684\\
14.3124651912943	0.062376384265729\\
15.1124651912943	0.0568469624860093\\
16.3124651912943	0.0432721019337379\\
17.9124651912943	0.0247411670000908\\
19.1124651912943	0.0140155597214502\\
20	0.00824503798100551\\
};
\addlegendentry{Simulation \eqref{eq:closed-loop-prediction-continuous}}

\addplot [color=mycolor1, dashed, line width=1.3pt]
  table[row sep=crcr]{%
0	0\\
0.159677448459238	-0.000272158485124407\\
0.188845799041001	-0.00366543166060751\\
0.226523016288059	-0.0120095675766407\\
0.278146833589389	-0.0275291769531343\\
0.342407851457516	-0.0505259135122991\\
0.411728815426226	-0.0830722153984134\\
0.503542050782386	-0.137591153218814\\
0.595355286138549	-0.202518414231719\\
0.7056945597813	-0.293824799597676\\
0.81937024917222	-0.40082096064787\\
0.957439664411741	-0.54566450885391\\
1.08806130642543	-0.694847006791324\\
1.21868294843911	-0.853153352138122\\
1.36601447292783	-1.03957762283478\\
1.6158554194513	-1.36582192344474\\
1.89517223029948	-1.73045264640694\\
2.02577375764173	-1.89620101821676\\
2.15637528498397	-2.05675652496496\\
2.32370331996993	-2.2527954942467\\
2.44805285349425	-2.39010272956844\\
2.57240238701857	-2.51923509914653\\
2.72639676579594	-2.66676276179421\\
2.87453732231836	-2.79473530851429\\
3.02267787884078	-2.90811166428704\\
3.19147902194516	-3.01871829367456\\
3.37160112596816	-3.11425038611982\\
3.53992509679302	-3.18212476734736\\
3.72143030003791	-3.23211437832754\\
3.87872268001913	-3.25623048564766\\
4.03601506000034	-3.26293275940877\\
4.20038072047503	-3.25187312232886\\
4.36199248855401	-3.22371291416095\\
4.50887168570948	-3.1840055468733\\
4.65320652411107	-3.13259407604922\\
4.82082614038504	-3.05840399657538\\
4.9904130971448	-2.96869996038117\\
5.12641643340964	-2.88717519587154\\
5.25369785694964	-2.80381718444332\\
5.41374181885854	-2.69018888496123\\
5.57378578076745	-2.5678539594042\\
5.71783810495326	-2.45143148356411\\
5.85469279896928	-2.33616254913442\\
6.02570707964141	-2.18684597370018\\
6.29332274181753	-1.94479373355858\\
7.03545757628943	-1.26015154074526\\
7.20461979092555	-1.10876822762921\\
7.37057713544851	-0.964080153418887\\
7.51573171712018	-0.841351858596362\\
7.66170196904499	-0.722020147767729\\
7.84925705106988	-0.575330581523122\\
7.99595241738585	-0.466328681151815\\
8.14264778370182	-0.362747165183677\\
8.33505690770545	-0.235489923816186\\
8.49800399436933	-0.135699235282736\\
8.66095108103321	-0.0435049902954709\\
8.8487706090792	0.0531853344092603\\
9.07714473306257	0.156831729577902\\
9.31583523035072	0.248777709038997\\
9.47449065762413	0.300781390086808\\
9.63314608489753	0.345763274272461\\
9.81485497062519	0.388858479783657\\
10.0153274904888	0.426298002127623\\
10.1814677971511	0.449801327110745\\
10.3220376718665	0.464671188524552\\
10.4845803464672	0.476322766306406\\
10.6584789417996	0.482558040360473\\
10.832377537132	0.482850385686699\\
10.990073418225	0.478451974637554\\
11.143969955408	0.470242216948154\\
11.3178852423084	0.456700485887183\\
11.5003259684037	0.438153055062504\\
11.6593642092114	0.41889907226853\\
11.8054935774679	0.399028227531232\\
11.9759951964137	0.373524769434336\\
12.176186473024	0.340904365452214\\
12.4808493325108	0.287633636419681\\
13.0066621306977	0.190735920812354\\
13.323209546288	0.132985337799898\\
13.7144098104163	0.0653251247255433\\
14.038628237431	0.0143987831198302\\
14.2211542568704	-0.01177397201565\\
14.4669505545527	-0.0440295355523688\\
14.7641090655035	-0.0776219632642174\\
15.0113357267956	-0.10113577925485\\
15.2427604174603	-0.119489918130668\\
15.474185108125	-0.134255381933475\\
15.6876693960562	-0.144665215238437\\
15.9131472629561	-0.152440199512686\\
16.2280286509626	-0.158553622308911\\
16.4467470977788	-0.159697161361496\\
16.8065195071292	-0.156602725741784\\
17.188550494616	-0.147621873176767\\
17.5603598926261	-0.134427597507926\\
17.9314249304145	-0.118110195650669\\
18.6719907623245	-0.080348012608372\\
19.4559782075161	-0.0398320411535877\\
20	-0.014836566844501\\
};
\addlegendentry{Simulation \eqref{eq:error-dynamics-delay}}

\end{axis}
\end{tikzpicture}%
\caption{Internal dynamic state $x_3$.}
\end{subfigure} \vspace{2mm}
\begin{subfigure}{\linewidth}
\input{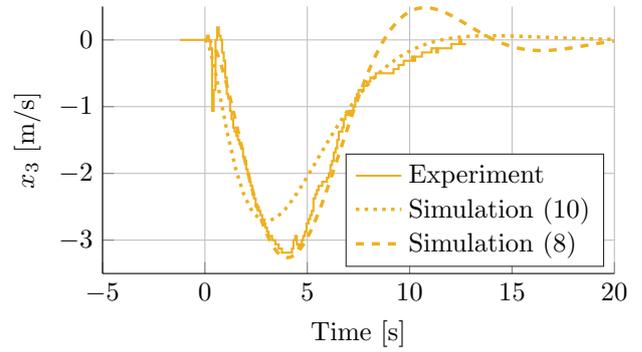}
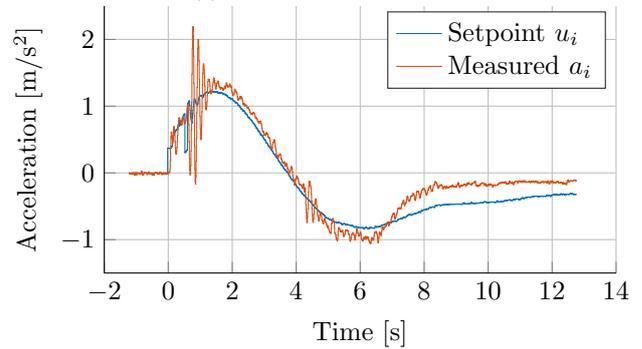
\caption{Acceleration $a_i$ subject to input $u_i$.}
\label{fig:acceleration-experiment-conventional}
\end{subfigure}
\caption{Measured experimental response of vehicle deploying conventional PD controller \eqref{eq:change-of-input}, compared to simulated response of \eqref{eq:error-dynamics-delay} and \eqref{eq:closed-loop-prediction-continuous}.} 
\label{fig:conventional-controller-experimental-response}
\end{center}
\end{figure}

Deploying the (conventional) controller \eqref{eq:change-of-input} on the experimental setup also nicely resembles the simulated closed-loop system of that respective controller as shown in \figref{fig:conventional-controller-experimental-response}. Similar observations can be made with respect to the match of the experimental results and simulation. For positive accelerations there is a good match, where for (small) negative accelerations the setpoint is again not reached by the vehicle at low velocities. This also explains the absence of an overshoot in the experimental results, as opposed to the simulated response. Additionally, a large dip in the error can be seen around 0.5 seconds in \figref{fig:conventional-controller-experimental-response-error}. At this point, the radar detected a ghost object causing the measured distance $d_i$ to be incorrect for a few samples. However, the overall response of the closed-loop system is still a good match on the time interval $t \in [0, 7]$ seconds and clearly indicates the difference in response of the prediction controller \eqref{eq:prediction-controller-discrete} and the controller \eqref{eq:change-of-input}. 

Summarizing, the experimental results confirm the prediction based controller indeed results in a closed-loop response according to \eqref{eq:closed-loop-prediction-continuous}, thereby effectively compensating the actuation delay of the experimental vehicle.

\section{Conclusion} \label{sec:conclusion}
A predictor feedback based CACC controller is presented that is able to effectively compensate the actuation delay of the ego vehicle. By using a discrete-time implementation of the controller, a finite dimensional representation of the controller is obtained that can be implemented in practice. By performing the stability analysis of the discrete-time closed-loop system, the effects of sampling times can be included to select appropriate controller gains. Finally, experiments with full-scale vehicles confirm the prediction-controller is able to successfully compensate the actuation delay in practice.

This work mostly focuses on eliminating the effects of the actuation delay of the ego vehicle on the closed-loop dynamics. However, topics such as controller tuning and string stability have not been addressed. Future work entails investigating the controller tuning and effects on performance metrics such as string stability characteristics, as well as deriving bounds on the errors. Furthermore, a better prediction of the preceding vehicle's acceleration during the dead-time of the ego vehicle can be included in the controller to further improve the performance. Finally, more extensive experiments should be performed where the initial-condition response is evaluated in a wider operating range, for instance during driving conditions at higher velocities instead of from standstill as in this paper.

\bibliography{references}        

\end{document}